# A new modeling approach to the safety evaluation of N-modular redundant computer systems in presence of imperfect maintenance

Francesco Flammini[1,2], Stefano Marrone[1,3], Nicola Mazzocca[2], Valeria Vittorini[2]

[1] ANSALDO STS - Ansaldo Segnalamento Ferroviario S.p.A.
Via Nuova delle Brecce 260, Naples, Italy
{flammini.francesco, marrone.stefano}@asf.ansaldo.it

[2] Università di Napoli "Federico II"
Dipartimento di Informatica e Sistemistica
Via Claudio 21, Naples, Italy
{frflammi, nicola.mazzocca, valeria.vittorini}@unina.it

[3] Seconda Università di Napoli
Dipartimento di Ingegneria dell'Informazione
Via Roma 29, Aversa (CE), Italy
stefano.marrone@unina2.it

**Abstract.** A large number of safety-critical control systems are based on N-modular redundant architectures, using majority voters on the outputs of independent computation units. In order to assess the compliance of these architectures with international safety standards, the frequency of hazardous failures must be analyzed by developing and solving proper formal models. Furthermore, the impact of maintenance faults has to be considered, since imperfect maintenance may degrade the safety integrity level of the system. In this paper we present both a failure model for voting architectures based on Bayesian Networks and a maintenance model based on Continuous Time Markov Chains, and we propose to combine them according to a compositional multiformalism modeling approach in order to analyze the impact of imperfect maintenance on the system safety. We also show how the proposed approach promotes the reuse and the interchange of models as well the interchange of solving tools.

**Keywords**

Safety, N-Modular Redundancy, Bayesian Networks, Imperfect Maintenance, Multiformalism Modeling

## 1. Introduction

Safety-critical computer systems (e.g. the ones used in hard real-time control applications) are often based on N-modular redundant architectures, using majority voters on the outputs of independent computation units. In order to assess the compliance of these architectures with international safety standards, the probability of the occurrence of unsafe events should be evaluated by developing and analyzing proper formal models. At this aim modeling languages for dependability evaluation such as Fault Trees and Stochastic Petri Nets have been widely used. A different approach is proposed in [16], where a Bayesian Network (BN) model is developed to evaluate the Mean Time Between Hazardous Events (MTBHE) of "2 out of 2" ("2oo2") voting architectures used in railway control applications in place of a Generalized Stochastic Petri Net (GSPN) model of the same system presented in [2]. The goal of that work was to point out that Bayesian Networks (more in general non state-based approaches) may be effectively used if the safety analysis can be performed by pure probabilistic techniques. This requires defining preconditions (i.e. constraints) or presumptions on vulnerability and failure mechanisms. Specifically, safe assumptions on fault/error latencies may be made on the voting architecture, provided that the result of the analysis is compliant with the safety requirements. In these cases, the BN model has several advantages over state-based models.

It is sometimes required by the safety assessors to evaluate the impact of maintenance faults (mainly due to human errors) on the safety integrity level of the system. Therefore, in this paper we extend the work presented in [16] to take into account imperfect maintenance. Since state-based approaches are usually needed when maintainability aspects must be explicitly addressed [14], here we propose a multiformalism approach which allows to develop two different independent models, a maintenance model and a failure-model, and combine them to analyze the impact of maintenance faults on the system integrity level. By separating the maintenance model and the failure model, it is possible to use different modeling languages, so that the failure model can be expressed by means of Bayesian Networks (if it is the case), and the maintenance model can be expressed by means of state-based formalisms. At the same time, the two models can be composed in order to obtain a multiformalism model allowing the two models to interact with each other. By properly defining the semantics of the composition and the interface of the models, the proposed approach promotes flexibility and model reuse (e.g. the failure model may be easily substituted by a state-based one whenever necessary, or the maintenance model may be substituted by another one if a more efficient solution technique or analysis tool is available).

This paper is organized as follows: Section 2 discusses some related works; Section 3 describes the reference architecture and the events that may cause hazardous failures; Section 4 synthesizes the model and the results presented in [16], Section 5 introduces imperfect maintenance and several maintenance models based on Continuous Time Markov Chains; Section 6 presents a multiformalism composed model accounting for imperfect maintenance and provides an example of evaluation of such model; finally, Section 7 contains conclusions and some hints about future work.

## 2. Related works

An introduction to the dependability of computer systems, including safety related aspects and a reference taxonomy, is provided by Avizienis at al. [3]. In the study reported in [2] the MTBHE of "2oo2" redundant architectures is evaluated by means of a GSPN model. In the same work it is shown how the result can be easily used for the safety evaluation of "2 out of 3" ("2oo3") architectures, which are widely employed in control applications and better known as Triple Modular Redundancy (TMR). The analysis of TMR systems ha been addressed by means of model-based studies on the hazardous failure rate using Markov Chains in the works of Hyunki et al. [19] and DeLong et al. [11], the latter focusing on safety related metrics.

As an alternative to the traditional Markov Chains models, Dynamic Fault Trees (DFT) [13] have also been introduced as a flexible and easy to use formalism for dependability modeling of TMR systems. DFT extends the Fault Tree formalism, which is only combinatorial, by adding state-based features. Repairable Fault Trees (RFT) [10] can be also used to model behavioral aspects of reliability when the focus is on system's maintenance [14]. In spite of the increased expressive power, both DFT and RFT feature the drawback of a lower solving efficiency, due to the required state-based analysis.

With the aim to overcome some of the efficiency limitations of the state-based formalisms, the application of Bayesian Networks to reliability modeling has been introduced by Portinale et al. [24]. A comparison of the efficiency of different (probabilistic) techniques for safety and dependability assessment can be found in reference [5], where it is underlined the tradeoff between expressive power and solving efficiency offered by BNs versus Fault Trees and GSPNs. Basing on the results of the aforementioned works, a study on multiformalism reliability evaluation of a complex railway control system based on Fault Trees and Bayesian Networks has been performed by the same authors of this paper [15]. The study applies to a real-world complex case-study the theoretical results provided in the previous works, highlighting some of the advantages of multiformalism modeling.

Imperfect maintenance of TMR systems is addressed in its general terms in [29], where both preventive and corrective (i.e. post failure) maintenance are considered. However, in such work it is assumed that corrective maintenance always renews the system and only preventive maintenance can be imperfect. Furthermore, the study is focused on system availability and does not consider safety related issues. Also the work reported in [18] focuses on preventive maintenance to avoid system deterioration. In this work Generalized Stochastic Petri Nets are used to represent and analyse the model. Even though related to safety-critical computer architectures, it seems that none of the existing works evaluates the impact of imperfect corrective maintenance (due to human or non-human factors) on system safety.

To the best of our knowledge, the work presented in this paper is the first study addressing multiformalism modeling in order to cope with imperfect maintenance in the safety analysis of voting architectures.

## 3. System description

Safety-critical systems must be compliant to international standards defining quantitative requirements on the Hazardous Failure Rate (HFR). One of these standards is the CENELEC 50129 [7], from which requirements of more specific safety specifications (e.g. ERTMS/ETCS [27]) are derived. Functional safety is assessed by means of qualitative approaches aimed at defining system functional safety requirements (e.g. Failure Mode & Effect Analysis, Hazard Analysis [12]) and allocating Safety Integrity Levels (SIL) to subsystems according to their criticality. More specifically, for the maximum level of safety criticality (SIL-4), Table A.1 of reference [7] (Section A.5.2) specifies the following quantitative requirement on the Tolerable Hazard Rate (THR): $10^{-9} \leq \text{THR} < 10^{-8}$. In order to fulfill such requirement, the usage of redundancy and majority voting techniques is widely adopted in implementing safety nucleuses. This reduces the chances of propagation of hardware faults till the application software layer and therefore the risk of catastrophic system failures.

A schematic representation of a TMR architecture is shown in Figure 1. TMR is a well known technique based on majority voting currently adopted to implement the safety nucleus of most critical control systems. It is based on three independent computing subsystems (or "units") whose output is voted according to a "2oo3" scheme. Exclusion logic units are external fail-safe devices used to shut-down a failed unit, that is the unit which is not agreeing with the remaining two. When a single unit fails it is excluded and the system continues to operate by relying on two units according to a "2oo2" voting scheme.

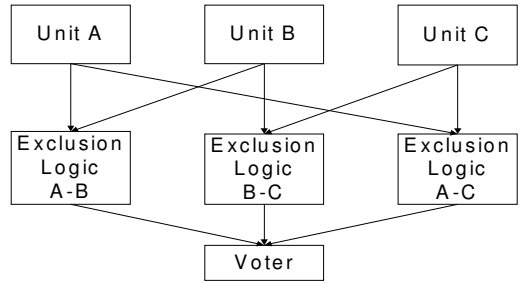

Figure 1. A Triple Modular Redundant architecture.

The complete independence of the three units is guaranteed by hardware fault-prevention and fault-tolerance techniques (e.g. galvanic isolation and fiber-optics based communication facilities). At the software level, redundancy codes, design diversity and consistency checks are employed, together with test and diagnostic processes. Furthermore, software errors are always considered as systematic in safety-critical systems and therefore they are not quantifiable, as also stated by the reference standard [7]. Thus, the dependability model of these architectures may not take into account common mode of failures. However, the modeling technique presented in this paper also allows to easily evaluate the impact of fault correlation on the hazardous failure rate, whenever required, though this aspect is out of the scope of this work.

The qualitative Fault Tree depicted in Figure 2 represents five paths of events leading to the occurrence of a hazardous failure, namely:

1. fault in the voter which affects the output;
2. simultaneously occurrence of a wrong output of one faulty unit and a voter failure in detecting unmatching outputs;
3. undetected invalid inputs to both units;
4. undetected invalid outputs from both units;
5. combined activation of two latent errors.

Given the real system architecture [2], paths 1 and 3 can be neglected. As for path 1, while in the logical scheme of Figure 1 the voter is represented as a separate unit, in most real implementations a distributed software voting is adopted. Therefore, voter failure is already considered in paths 4 and 5. As for path 3, the correctness of inputs is assured by external mechanisms (e.g. redundancy codes) which are not in the scope of this work.

A more dynamic representation of the behavior of a "2oo2" system which is a step closer to the real world is provided by the flowchart in Figure 3, which also takes into accounts the dynamic aspects related to the error propagation and the interaction between the two units. Errors generated by transient (i.e. software) faults are assumed to remain latent for a period 1/Lt before eventually activating and generating a unit failure (i.e. an incorrect output), when they are not detected by the internal diagnostic checks. System behaviour with permanent (i.e. hardware) faults differ from the one described for transient faults in the following points: 1) the average latency, which is 1/Lp; 2) the fact that a (small) class of hardware failures can be non diagnosable (that is, they always activate); 3) error generation from a permanent fault is continuous, until system shutdown. Due to the redundancy provided by system architecture, a single unit failure does not generate a hazardous output unless a failure occurs also in the exclusion logic or both units produce the same wrong output (a very rare event).

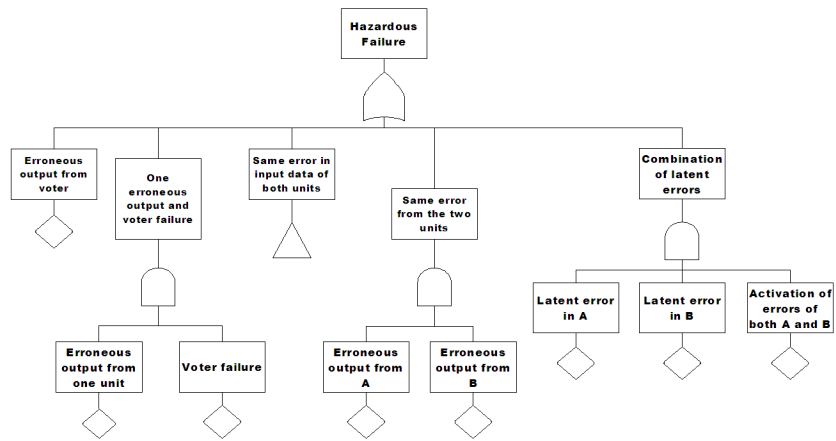

Figure 2. The Fault Tree for the hazardous failure of the "2oo2" architecture.

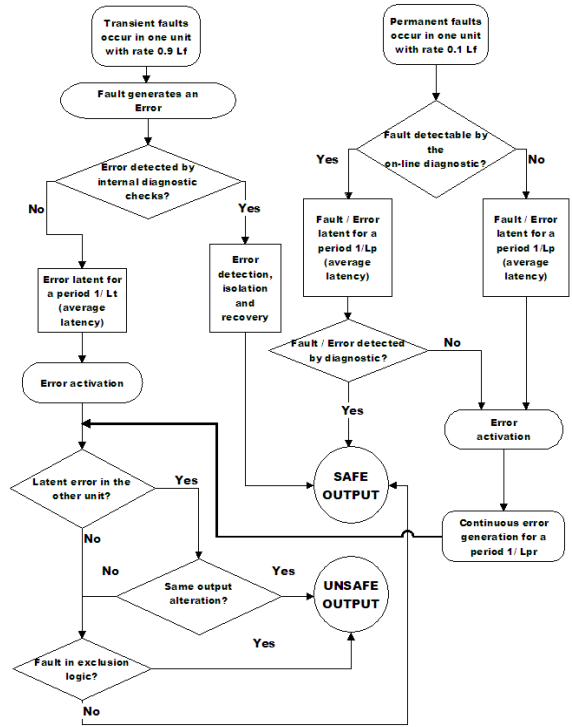

Figure 3. The hazardous failure flow-chart.

## 4. Failure model

In this Section we describe the failure model for the MTBHE evaluation formerly introduced in [16]. It is obtained by translating the flow-chart shown in Figure 3 into a Bayesian Network. We believe that Bayesian Networks can be advantageously employed to address the hazardous failure modeling of "2oo2" architectures, since the resulting model is:

- easier to manage and hence less error prone when performing maintenance (as it will be demonstrated in the rest of this section);
- quicker to evaluate (as quantitatively demonstrated in reference [5]).

Furthermore, the specific features of BN also allow for useful automatic analyses, like sensitivity to findings and most probable explanations, which are supported by most solving tools (e.g. Netica [23]).

In fact, the translation of the hazardous failure flow-chart into the Bayesian Network depicted in Figure 4 is quite straightforward. Events represented in the flow-chart are associated to random variables in the BN representing their probability of occurrence, while arcs of the flow-chart are transformed in corresponding arcs of the BN which are assigned conditional probability values associated to the related cause-effect relationships. This is done separately for the two units and the feasibility of the translation process is due to the absence of loops in the resulting network (BN are acyclic graphs).

The main difference between the BN model and the GSPN model described in [2] is that the BN model is purely probabilistic while the GSPN model explicitly considers timing (and therefore state-related) properties. When timing aspects must be introduced in a probabilistic model, one possible way is to make some safe assumptions, i.e. by considering the worst-case conditions in the choice of the net parameters. For instance, if there is a possibility for the on-line diagnostic checks to detect latent errors, the safe assumption consists in neglecting such possibility: if the evaluation result is compliant with system safety requirements, then we can be satisfied with such result; otherwise, we need to refine the model. In the example, the refinement can be performed by means of an "off-line" analysis and prediction of the detection probability of the diagnostic checks during the reference time interval, which is 1 hour in case of the MTBHE model; the result of the analysis is finally embedded in the model as a plain probability.

The Conditional Probability Table (CPT) for the random variable modeling the occurrence of the UNSAFE_OUTPUT event is shown in Table 1.

The parameters and the probabilities used to populate the model have been listed in Table 2. The reported values are fixed in this study, since they are related to hardware fault probabilities, design choices and measured data.

The "2oo3" failure model can be derived from the "2oo2" model:

$$MTBHE_{2oo3} = 1/3 \cdot MTBHE_{2oo2}$$

In fact, the behavior of a "2oo3" system can be described by three "2oo2" systems: units A AND B, units B AND C, units A AND C.

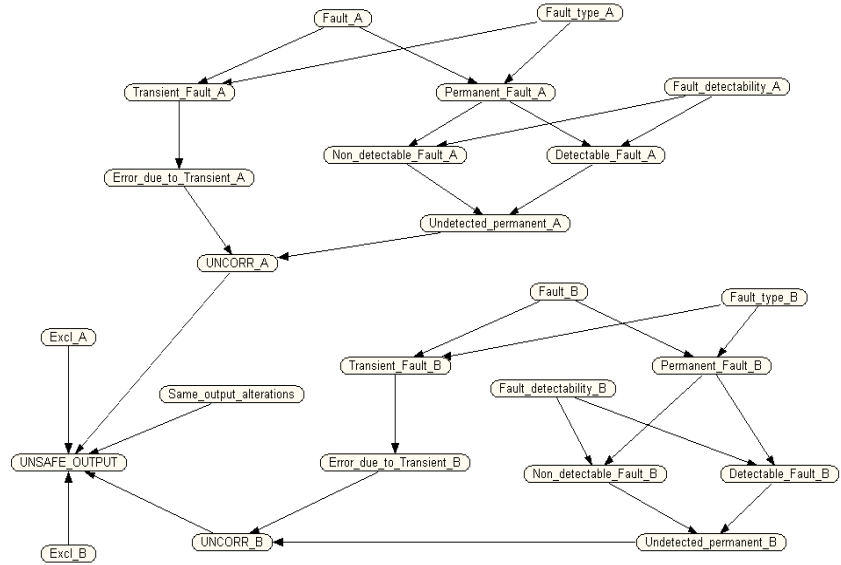

Figure 4. The Bayesian Network model of the "2oo2" computer architecture.

| True | False | Same Output Alt. | UNCORR_A | UNCORR_B | Excl_A | Excl_B |
|---|---|---|---|---|---|---|
| **1** | **0** | True | True | True | True | True |
| | | | | | True | False |
| | | | | | False | True |
| | | | | | False | False |
| | | | | False | True | True |
| | | | | | True | False |
| **0** | **1** | | | | False | True |
| **0** | **1** | | | | False | False |
| **1** | **0** | | False | True | True | True |
| **0** | **1** | | | | True | False |
| **1** | **0** | | | | False | True |
| | | | | | False | False |
| **0** | **1** | | | False | True | True |
| | | | | | True | False |
| | | | | | False | True |
| | | | | | False | False |
| **1** | **0** | False | True | True | True | True |
| **1** | **0** | | | | True | False |
| **1** | **0** | | | | False | True |
| **0** | **1** | | | | False | False |
| **1** | **0** | | | False | True | True |
| **1** | **0** | | | | True | False |
| **0** | **1** | | | | False | True |
| **0** | **1** | | | | False | False |
| **1** | **0** | | False | True | True | True |
| **0** | **1** | | | | True | False |
| **1** | **0** | | | | False | True |
| **0** | **1** | | | | False | False |
| **0** | **1** | | | False | True | True |
| | | | | | True | False |
| | | | | | False | True |
| | | | | | False | False |

Table 1. The Conditional Probability Table for the UNSAFE_OUTPUT event.

The model has been evaluated in order to demonstrate system compliance with safety requirements. For instance, reference [27] reports the safety requirements of a modern railway control systems (i.e. ERTMS/ETCS [26]). In this document, the THR requirement of $2 \cdot 10^{-9}$ hazardous failures per hour is apportioned on the ground subsystem for one half (i.e. $10^{-9}$ HF/h), the MTBHE being obviously the inverse of the THR. Generally speaking, values in the order of $10^{-9}$ are common when defining quantitative safety requirements.

The tool we used for model construction and evaluation is Netica [23]. The results of the analysis are reported in Table 2, where the last row (shaded in grey) highlights the main result, consisting in the evaluated Hazard Rate (HR) of the system.

Quite obviously:

$\quad$ MTBHE$_{2oo3}$ = $1 / (3 \cdot$ HR$_{2oo2}) = 6.9362 \cdot 10^{11}$ h

which is largely compliant to system safety requirements, also considered the other possible causes of hazardous failures. Compared with the data provided in reference [2], these results have served to validate our model.

Figure 5 reports the results of an "a posteriori" probability evaluation (with the UNSAFE_OUTPUT event value fixed to TRUE). Such results allows engineers to understand which is the relative contribution of each factor to system safety without performing a number of sensitivity analyses: more "critical" events are simply the ones with the higher "a posteriori" probabilities. For instance, transient faults give a higher contribution (68.4%) to hazardous failures with respect to permanent faults (31.6%), while only the 7.6% of hazardous failures are due to non detectable faults. These results suggest that if an effort has to be performed in order to increase system safety, then it should be allocated to transient fault prevention and tolerance. Checking for symmetry in the results is also useful to validate the model.

We recall that Bayesian Networks only allows for a probabilistic approach, with safe assumptions on fault/error latencies. Safe assumptions can be very effective in simplifying models; however, they usually take to a worsening of the results. As the analysis is aimed at fulfilling a safety-requirement given by a reference standard, the assumptions can be considered acceptable as long as the result is below the specified threshold, otherwise, the model has to be refined in order to achieve better results or a different modeling formalism must be employed.

The results obtained in this Section do not account for imperfect maintenance (it is assumed that all the repair interventions are performed correctly), which is addressed in next section.

| STOCHASTIC VAR. | MEANING | VALUE | PROBABILITY |
|---|---|---|---|
| Detectable_Fault_A | A detectable fault occurring in Unit A | True | $1.5 \cdot 10^{-6}$ |
| Detectable_Fault_B | A detectable fault occurring in Unit B | True | $1.5 \cdot 10^{-6}$ |
| Error_due_to_Transient_A | An error due to a transient fault occurring in Unit A | True | $1.5 \cdot 10^{-6}$ |
| Error_due_to_Transient_B | An error due to a transient fault occurring in Unit B | True | $1.5 \cdot 10^{-6}$ |
| Excl_A | A failure occurring in the Exclusion Logic of Unit A | True | $10^{-10}$ |
| Excl_B | A failure occurring in the Exclusion Logic of Unit B | True | $10^{-10}$ |
| Fault_A | A fault occurring in Unit A | True | $1.6666 \cdot 10^{-5}$ |
| Fault_B | A fault occurring in Unit B | True | $1.6666 \cdot 10^{-5}$ |
| Fault_detectability_A | The possibility of detecting a fault occurring in Unit A | Detectable | 0.9 |
| Fault_detectability_B | The possibility of detecting a fault occurring in Unit B | Detectable | 0.9 |
| Fault_type_A | The fault type of Unit A (Transient or Permanent) | Transient | 0.9 |
| Fault_type_B | The fault type of Unit B (Transient or Permanent) | Transient | 0.9 |
| Non_detectable_Fault_A | A non detectable fault occurring in Unit A | True | $1.6666 \cdot 10^{-7}$ |
| Non_detectable_Fault_B | A non detectable fault occurring in Unit B | True | $1.6666 \cdot 10^{-7}$ |
| Permanent_Fault_A | A permanent fault occurring in Unit A | True | $1.6666 \cdot 10^{-6}$ |
| Permanent_Fault_B | A permanent fault occurring in Unit B | True | $1.6666 \cdot 10^{-6}$ |
| Same_output_alterations | The two units produce the same modification of their output | True | 0.1 |
| Transient_Fault_A | A transient fault occurring in Unit A | True | $1.5 \cdot 10^{-5}$ |
| Transient_Fault_B | A transient fault occurring in Unit B | True | $1.5 \cdot 10^{-5}$ |
| UNCORR_A | An uncorrect output occurring in Unit A | True | $2.1912 \cdot 10^{-6}$ |
| UNCORR_B | An uncorrect output occurring in Unit B | True | $2.1912 \cdot 10^{-6}$ |
| Undetected_permanent_A | An undetected permanent fault occurring in Unit A | True | $6.9164 \cdot 10^{-7}$ |
| Undetected_permanent_B | An undetected permanent fault occurring in Unit B | True | $6.9164 \cdot 10^{-7}$ |
| **UNSAFE_OUTPUT** | **An unsafe output occurring in the system** | **True** | **$4.8056 \cdot 10^{-13}$** |

Table 2. Bayesian Network parameters and results.

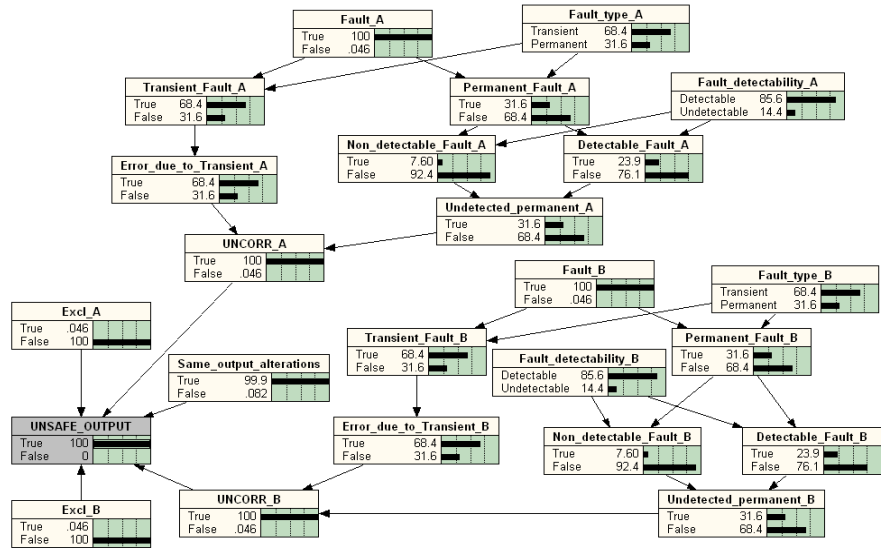

Figure 5. Model evaluation by means of "a posteriori" probabilities.

## 5. Maintenance model

The aim of this Section is to evaluate the impact of imperfect maintenance on system safety. To this aim, Continuous Time Markov Chains (CTMC) are employed.

In general, non state-based approaches (e.g. Bayesian Networks) feature the advantage of an improved efficiency over the state-based approaches (e.g. Markov Chains, Timed Automata, Stochastic Petri Nets) for stochastic analysis of dependability properties, with combinatorial approaches being the most efficient (e.g. Fault Trees, Reliability Block Diagrams) [5]. However, state-based approaches are usually needed when maintainability aspects must be explicitly taken into account [14].

From the above considerations, it is clear that when modeling safety related aspects a trade-off must be achieved between modeling expressiveness, solving efficiency and ease of use of modeling formalisms, using the most straightforward, readable, maintainable and quick to evaluate model which is able to fulfill the evaluation objectives. A similar consideration should be done when deciding the level of abstraction of models. This is particularly meaningful when the MTBHE model needs to be enriched with imperfect maintenance modeling, which is a real industrial need in the safety assessment process.

Imperfect maintenance modeling must take into account:

- the missed substitution of the faulty section after a possibly non diagnosable permanent hardware fault;
- the occurrence of permanent hardware faults during system shut-down;
- the unwilled system restart after a loss and return of power.

Hence, a state-based model is required and it can be built at different levels of detail, from four to eight states, as shown in Figure 6. In order to consider only four states, the modeler needs to make safe assumptions. More specifically, all permanent faults must be considered as non diagnosable. This obviously leads to an underestimation of the MTBHE. The other common assumptions are:

1) The unsafe failure states give a negligible contribution;

2) The most severe failure states can include less severe ones (e.g. the state with non diagnosable permanent faults does not exclude the presence of diagnosable permanent faults).

We will describe the three models starting from the simplest and adding details for the more complex ones. The states of all three models are described in Table 3.

| State | Extended Name | Description |
|---|---|---|
| $S_0$ | SYSTEM UP WITH NO UNDIAGNOSABLE PERMANENT FAULTS | The system is operating normally, without latent non diagnosable permanent faults |
| $S_1$ | SYSTEM SHUTDOWN WITH NO PERMANENT FAULTS (POWERED) | The system is in the safe shutdown state with no permanent faults accumulated |
| $S_2$ | SYSTEM SHUTDOWN WITH A NON DIAGNOSABLE PERMANENT FAULT (POWERED) | The system is in the shutdown state after a non diagnosable permanent faults has occurred |
| $S_3$ | SYSTEM UP WITH A NON DIAGNOSABLE PERMANENT FAULT | The system is in an unsafe state with permanent faults which will eventually activate |
| $S_4$ | SYSTEM NOT POWERED WITH A NON DIAGNOSABLE PERMANENT FAULT | The system is unpowered after at least a non diagnosable permanent fault has been accumulated |
| $S_0'$ | SYSTEM UP WITH NO PERMANENT FAULTS | The system is operating normally without latent permanent faults |
| $S_0''$ | SYSTEM UP WITH A DIAGNOSABLE PERMANENT FAULT | The system is operating with at least a permanent fault which has not been detected yet |
| $S_5$ | SYSTEM SHUTDOWN WITH A DIAGNOSABLE PERMANENT FAULT (POWERED) | The same of $S_2$ for diagnosable faults |
| $S_6$ | SYSTEM NOT POWERED WITH A DIAGNOSABLE PERMANENT FAULT | The same of $S_4$ for diagnosable faults |

Table 3. States of the imperfect maintenance model.

The simplest model is depicted in Figure 6 (c) and it is composed of four states. The initial state is $S_0$, representing normal system operation. In case of a safe shutdown, the system moves to state $S_1$, while if a non diagnosable permanent fault occurs, the next state will be $S_3$ (representing a hazardous state). Starting from $S_1$, it is possible to reach $S_2$ if a non diagnosable error occurs or to return back to $S_0$ whereas a maintenance intervention has been performed to restore the system. System can move from $S_2$ to $S_0$ in case of correct maintenance or to $S_3$ in case of loss and return of power. Finally, moving from $S_3$ to $S_2$ is allowed by a safe shutdown (e.g. caused by another diagnosed error).

The "intermediate" model of Figure 6 (b) introduces a new state ($S_4$) which can be reached from $S_3$ and $S_2$ whereas a loss of power occurs (when the power is restored, system returns to $S_3$).

The most complex model of Figure 6 (a) adds the following features:

- partitioning of $S_0$ state in two states ($S_0'$ and $S_0''$) which better highlight the effect of error detection latency on diagnosable faults;
- introduction of two states $S_5$ and $S_6$ modeling the same dynamics of $S_2$ and $S_3$ for diagnosable permanent faults.

As an aside, please note that the transitions between states can be classified into two main categories: in the first we can list the ones related to inner and non-human factors (e.g. system shutdowns in case of failures); the second comprises outer and human factors (e.g. incorrect repairs or ungoverned start-ups after a power restore).

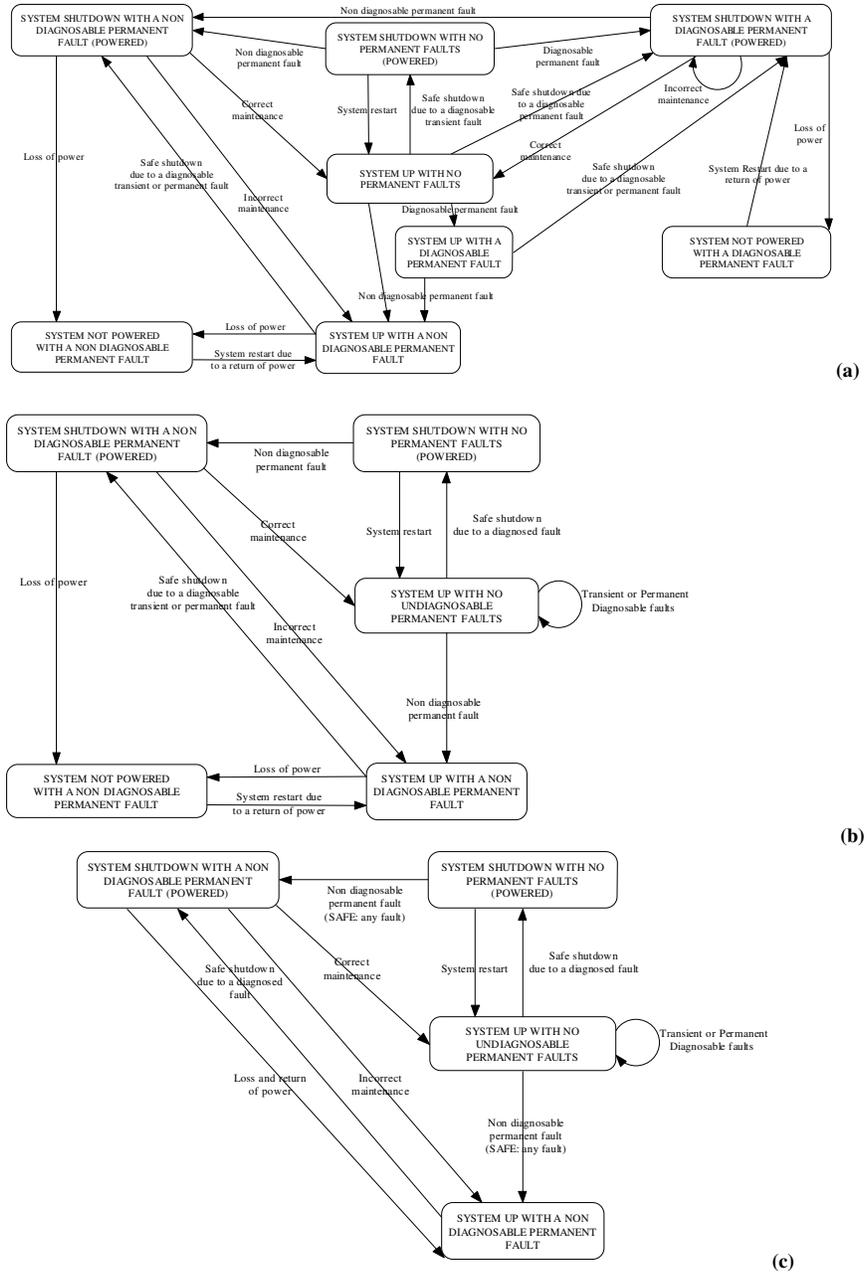

Figure 6. State-based maintenance models at different levels of detail.

## 6. A multiformalism model combining failure and maintenance models

The failure and maintenance models presented in the previous Sections may be developed and updated independently, but they can be composed and solved jointly in order to describe and analyze the real working system in presence of imperfect maintenance.

To this aim a multiformalism approach is needed since the two models are described by different modeling languages. The use of multiformalism techniques is very appealing in modeling complex systems as they allow operating on each aspect of interest by using the most appropriate modeling language and the best available solution technique [25]. While a single highly expressive formalism (e.g. GSPN) can always be employed in order to design the entire model, this would imply a poorly manageable model, which is more difficult to understand and validate, and harder to solve as a whole, due to the state-space explosion phenomenon.

Here we combine the failure and maintenance models applying the OsMoSys Modeling Methodology (OMM) [28], which is based on object orientation and metamodeling concepts. According to OMM, each model is an instance of its Model Class. A Model Class defines the structure of a model expressed by a specified formalism, the parameters that must be provided when instancing objects, and the interface of the model, that is the elements of the model used to interact with other models. The interaction is defined and implemented by means of proper composition operators. Hence, in the following the failure model and the maintenance model are two Model Classes (FM and MM respectively). Let us assume that FM is the BN model described in Section 4 and MM is the model of Figure 6 (b); their parameters are listed in the following.

FM MODEL CLASS:

- $PAR_1$: the fault probability of a single unit;
- $PAR_2$: the ratio of the probability of non diagnosable faults (in both units);
- $PAR_3$: the probability of two simultaneous faults producing identical outputs;

MM MODEL CLASS:

- $PAR_4$: probability of occurrence of an error in one of the two units (leading to system safe shutdown);
- $PAR_5$: UNSAFE_OUTPUT (the HFR of the "2oo2" system);
- $PAR_6$: inverse of Mean Time to Repair (MTTR);
- $PAR_7$: ratio of wrong maintenance interventions;
- $PAR_8$: inverse of Mean Time Between Failures (MTBF) of the power line;

- PAR$_9$: inverse of the Mean Time To Restore (MTTRS) of the power line.

In the following we clarify the relationships between the above parameters and the CTMC models:

- the safe shutdown parameter, that is the probability of the transition from $S_0$ to $S_1$ and from $S_3$ to $S_2$, is calculated as 2*PAR$_4$-PAR$_5$. This is due to the fact that a safe shutdown occurs every time at least one unit fails (PAR$_4$ for two units) except the time when an unsafe shutdown occurs (PAR$_5$);

- the unsafe shutdown parameter (transition probability from $S_1$ to $S_2$ and from $S_0$ to $S_3$) is simply PAR$_5$;

- the system restart parameter (transition from $S_1$ to $S_0$) is PAR$_6$;

- the incorrect maintenance parameter (transition from $S_2$ to $S_0$) is PAR$_6$ * PAR$_7$, while the correct one (transition from $S_2$ to $S_3$) is (1 - PAR$_7$) * PAR$_6$;

- the transition to $S_4$, representing the loss of power, has probability PAR$_8$;

- the transition from $S_3$ to $S_4$, representing the spontaneous restore of power, has probability PAR$_9$;

- the output of the model is the steady-state probability of $S_3$ which is represented by PAR$_{10}$.

In order to perform the analysis, one object $\phi$ of FM and one object $\mu$ of MM must be instantiated specifying the values of their parameters. Nevertheless, PAR$_4$ and PAR$_5$ may be obtained by solving a failure model. Hence, $\phi$ can be instantiated, but $\mu$ is fully instantiated only after the results of the failure model are available. This is a simple example of sequential composition, in which the interfaces of FM and MM contain the parameters PAR$_4$ and PAR$_5$, (output and input parameters, respectively). The composition operator used to combine the two models specifies that $\phi$ has to be solved and that the results represented by PAR$_4$ and PAR$_5$ must be used to instantiate $\mu$. This situation is depicted in Figure 7, in which PAR$_{10}$ is the steady state probability of the system being in the state "UP WITH UNDIAGNOSABLE PERMANENT FAULTS", which is obtained by solving $\mu$.

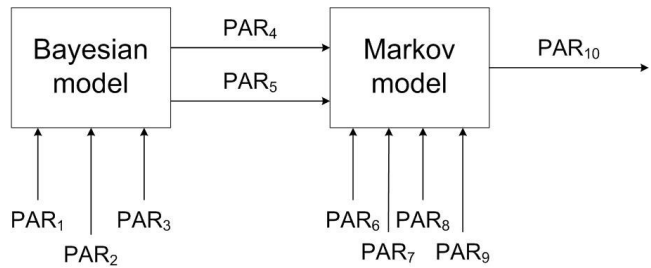

Figure 7. Composition of failure and maintenance models.

If we instantiate $\phi$ with the values:

$$PAR_1 = 1.666 \cdot 10^{-5}, PAR_2 = 10^{-1}, PAR_3 = 10^{-1}$$

the solution of the failure model produces the following results:

$$PAR_4 = 2.19 \cdot 10^{-6}, PAR_5 = 4.8 \cdot 10^{-13}.$$

Assuming that:

$$PAR_6 = 1, PAR_7 = 10^{-2}, PAR_8 = 10^{-4}, PAR_9 = 3$$

and solving the maintenance model $\mu$, we obtain that the HFR of the "2oo3" is $3.33 \cdot 10^{-7}$, which is greater than the required $10^{-9}$.

Sensitivity analyses can help to understand how to achieve better results. For example:

1) reducing $PAR_2$ by a factor of 10, there are negligible variations of the results;

2) reducing $PAR_1$ by a factor of 10, $PAR_4$ is reduced by a factor of 10 and $PAR_5$ is reduced by a factor of 100;

3) reducing $PAR_3$ by a factor of 10, $PAR_4$ remains unchanged while $PAR_5$ is reduced by a factor of 10.

The results of the second analysis highlight that the model is not much sensible to variations of $PAR_6$, $PAR_7$, $PAR_8$ and $PAR_9$, while it is very sensible to variations of $PAR_4$ and $PAR_5$. In particular, acceptable results can be achieved only if $PAR_4$ is high and $PAR_5$ is low.

Considering the following parameters:

$$PAR_1 = 10^{-5}, PAR_2 = 10^{-1}, PAR_3 = 3 \cdot 10^{-4}$$

the analysis of the BN model leads to the following results:

$$PAR_4 = 1.3 \cdot 10^{-6}, PAR_5 = 7.81 \cdot 10^{-16}$$

and the analysis of the Markov model leads to:

$$HFR_{2oo3} = 9.1 \cdot 10^{-10}.$$

This is an acceptable result, considered that the choice $PAR_3 = 3 \cdot 10^{-4}$ is justified by the fact that the probability of identical erroneous outputs is actually much lower [9].

More sophisticated compositions may be addressed by adopting a multiformalism modeling approach. Figure 8 shows a general scheme where the model interfaces and the composition operator are explicitly represented. As aforementioned, the choice of the models is related to the objectives of the evaluation.

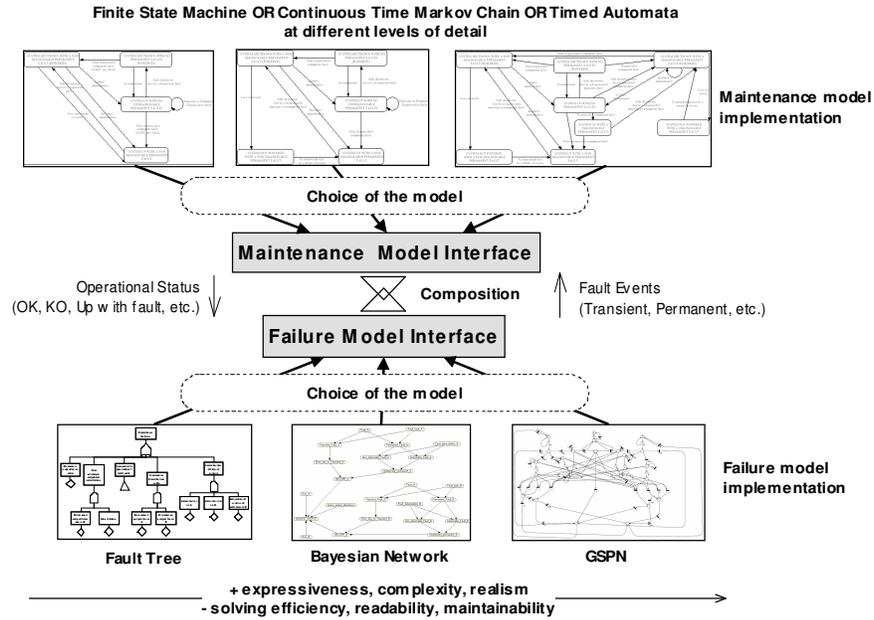

Figure 8. Multiformalism multilevel modeling.

Indeed, Figure 8 provides a generalization of the case-study analyzed above: the adopted multiformalism methodology provides the mechanisms by which multiple Model Classes may implement the same interface. This makes it possible to easily change, for example, the BN model with a GSPN or a FT version of the failure model, if needed, provided that they implement the same interface (in the specific case, if the parameters $PAR_4$ e $PAR_5$ can be obtained by solving the GSPN or the FT models). The same holds for the CTMC maintenance model: Figure 8 highlights the possibility of choice among the three models described in Section 5. The possibility of model reuse and the flexibility of analysis provided by a multiformalism approach reveals to be very useful when customizations of the failure or maintenance models are needed (e.g. to fit different installations of the system). Furthermore, it could happen that the maintenance model is provided by the customer, who manages the repair policies, while the failure model is managed by the supplier, who knows the details of the system architecture.

Operators are used to connect submodels and carry information about the composition semantics. OMM distinguishes two kinds of operators: operators implying model manipulation, and operators which require the evaluation of results from submodels in order to instantiate or modify other submodels. The operator used above to combine the failure and the maintenance models belongs to the second category. Nevertheless different composition semantics (i.e. different integration policies) between the failure and the maintenance models are possible. For example, the two model classes could be coupled in a ways such that if in the failure model the shutdown event is activated, then the maintenance model switches to the system shutdown state, and when the failure model is in system shutdown state, only permanent hardware faults can happen in the failure model. The overall composed model must then be solved as a whole. Again, if the result of the evaluation is compliant with the requirement, then model choice can be considered as satisfactory, otherwise more refined models/formalisms must be adopted (a priority based on readability or solving efficiency could be defined in order to guide the choice of the sub-model to be refined). If the safety requirement is not demonstrated to be fulfilled even with the most complicated model, then some of the design or maintenance parameters need to be changed.

The analysis of complex OMM models can be automated by means of the OsMoSys Multisolution Framework (OMF) [17]. The OMF has been developed to provide the support needed to a loosely coupled cooperation among heterogeneous analysis techniques and tools. In other words, it automates the tasks that must be performed to solve complex multiformalism models. "Multisolution" in OsMoSys means solving a composed model according to a well defined solution process. The solution process usually involves the execution of more solution or analysis tools (solvers). The execution order of the solvers and the data dependencies among them are defined by the solution process, which is described by means of a workflow language.

## 7. Conclusions

In this paper we have introduced a new approach to model N-modular redundant computer systems used in safety-critical control applications. The results we have obtained support the use of different formalisms for an easy and effective representation of the hazardous failure model. In particular, the issue of evaluating the impact of imperfect maintenance on system safety has been addressed by solving a multiformalism compositional model including a Bayesian Network failure model and a Continuous Time Markov Chain maintenance model. The mutiformalism approach could appear more complex with respect to one based on a single formalism. However, our experience has demonstrated that many classes of modeling problems (like the one described in this paper) are far better managed using "divide et impera" approaches, based on modular and incremental techniques (see e.g. [30]). A user-friendly support framework (OMF [17]) has been engineered to the scope, and it is currently under development, with the aim of hiding to the modeler the complexity of the solving process.

The choice of modeling formalism we have performed seems to fit well the case-study presented in this paper. However, it is always useful to explore further combinations of formalisms in order to enhance efficiency and ease of modeling when dealing with the same or different problems. For instance, basing on different assumptions, it is possible to employ Repairable Fault Trees [10] and/or Dynamic Bayesian Networks [21] (allowing to explicitly model state-base aspects). Both formalisms implement an implicit multiformalism paradigm which can be managed in the OsMoSys framework. Implicit multiformalism allows solving models using different formalisms while presenting to the modeler a single and possibly straightforward formalism. Furthermore, the solving process can be such to optimize efficiency. In the case of RFT, this is performed by using combinatorial techniques (i.e. Fault Tree) on non repairable tree branches and state-based techniques (i.e. GSPN) on repairable ones. Another way to improve efficiency is to recognize and exploit model symmetries (e.g. tree branches sharing the same structure) and "fold" them by means of a colored class of Petri Nets, e.g. Stochastic Well-formed Nets [8].

As mentioned in Section 6, a cohesive model which is more adherent to reality allows for less conservative assumptions, with the result of a better optimization of design parameters. Therefore, the OMM [17] is going to be extended with new and more sophisticated composition operators, allowing for a strictly coupled interaction between models (based on state or action sharing),